%% file: main.tex
\documentclass[%
aip,
 reprint,
 amsmath,amssymb,
 aps,
floatfix,
]{revtex4-1}

\usepackage{bm}
\usepackage{bbm}
\usepackage{dsfont}
\usepackage{graphicx}
\usepackage{mathtools}
\usepackage{array}
\usepackage{tikz}
\usepackage{color}
\usepackage{float}
\usepackage{xspace}
\usepackage{bm}
\usepackage{comment}
\usepackage{nicefrac}
\usepackage{amsmath}

\begin{document}

\preprint{APS/123-QED}

\title{Constant-adiabaticity ultralow magnetic field manipulations of parahydrogen-induced polarization: application to an AA$'$X spin system}

\author{Bogdan A. Rodin}
 \affiliation{International Tomography Center SB RAS, Novosibirsk, Russia}
 \affiliation{Novosibirsk State University, Novosibirsk, Russia}
\author{James Eills*}
\affiliation{Johannes-Gutenberg University, Mainz 55099, Germany}
\affiliation{Helmholtz Institute Mainz, GSI Helmholtzzentrum f{\"u}r Schwerionenforschung, 55128 Mainz, Germany}
\author{Rom\'{a}n Picazo-Frutos}
\affiliation{Johannes-Gutenberg University, Mainz 55099, Germany}
\affiliation{Helmholtz Institute Mainz, GSI Helmholtzzentrum f{\"u}r Schwerionenforschung, 55128 Mainz, Germany}
\author{Kirill F. Sheberstov}
\affiliation{Johannes-Gutenberg University, Mainz 55099, Germany}
\affiliation{Helmholtz Institute Mainz, GSI Helmholtzzentrum f{\"u}r Schwerionenforschung, 55128 Mainz, Germany}
\author{Dmitry Budker}
\affiliation{Johannes-Gutenberg University, Mainz 55099, Germany}
\affiliation{Helmholtz Institute Mainz, GSI Helmholtzzentrum f{\"u}r Schwerionenforschung, 55128 Mainz, Germany}
\affiliation{Department of Physics, University of California, Berkeley, CA 94720-7300}
\author{Konstantin L. Ivanov*}%
 \affiliation{International Tomography Center SB RAS, Novosibirsk, Russia}
 \affiliation{Novosibirsk State University, Novosibirsk, Russia}

\date{\today}

\input symbols/main.tex
\input ReferenceSets.tex

\begin{abstract}

The field of magnetic resonance imaging with hyperpolarized contrast agents is rapidly expanding, and parahydrogen-induced polarization (PHIP) is emerging as an inexpensive and easy-to-implement method for generating the required hyperpolarized biomolecules. Hydrogenative PHIP delivers hyperpolarized proton spin order to a substrate via chemical addition of H\textsubscript{2} in the spin-singlet state, but prior to imaging it is typically necessary to transfer the proton polarization to a heteronucleus (usually \textsuperscript{13}C) in the molecule. Adiabatic ultralow magnetic field manipulations can be used to induce the polarization transfer, but this is necessarily a slow process, which is undesirable since the spins continually relax back to thermal equilibrium. Here we demonstrate constant-adiabaticity field cycling and field sweeping for optimal polarization transfer on a model AA$'$X spin system, [1-\textsuperscript{13}C]fumarate. We introduce a method for calculating constant-adiabaticity magnetic field ramps and demonstrate that they enable much faster spin-order conversion as compared to linear ramps used before. The present method can thus be utilized to manipulate nonthermal order in heteronuclear spin systems.

\end{abstract}

\maketitle

\section{Introduction}
Parahydrogen induced polarization (PHIP)~\cite{natterer1997,duckett1999} is a widely used method to enhance NMR signals. The source of nonthermal spin order in PHIP experiments is the singlet order of parahydrogen ({\it p}H$_2$, molecular hydrogen in the nuclear spin-singlet state). Although {\it p}H$_2$ does not have a magnetic moment and is thus NMR-silent, upon symmetry breaking (i.e. by rendering the two protons chemically or magnetically inequivalent) the nonthermal singlet order can be converted into observable NMR signals, which are strongly enhanced compared to those under equilibrium conditions. The first step for hydrogenative PHIP is a catalytic hydrogenation reaction (addition of H$_2$ to a suitable substrate, usually one with an unsaturated \mbox{C-C} bond). When the two {\it p}H$_2$-nascent protons occupy inequivalent positions in the reaction product the symmetry is broken, and NMR signal enhancements can be obtained. The magnetic interaction that induces symmetry breaking is typically a chemical shift difference, or inequivalent \textit{J}-couplings to a third nucleus.

A common step in PHIP is transferring nonthermal spin order from the source spins -- here the {\it p}H$_2$-nascent protons -- to target spins of choice, which are more suitable for NMR detection for various reasons (longer relaxation times, higher spectral resolution, lower background signals). A number of methods have been developed to transfer the {\it p}H$_2$ spin order to various heteronuclei, via rf pulse methods at high field~\cite{goldman2006, kadlecek2010, baer2012, pravdivtsev2014, eills2017, stevanato2017a, stevanato2017, schmidt2017, korchak2018a, kozinenko2019}, or through coherent spin mixing under zero- to ultralow- field (ZULF) NMR conditions~\cite{stephan2002, johannesson2004, goldman2005, cavallari2015, kuhn2006, pravdivtsev2013c, eills2019}. In the ZULF regime, Larmor frequencies are small, and nuclear spins belonging to different isotopic species become ``strongly coupled'' -- that is the difference in Larmor frequencies becomes comparable to the spin-spin couplings. Under these conditions, coherent exchange of polarization among the spins becomes possible.

A number of polarization-transfer techniques exploiting ultra-low magnetic field manipulations have been developed, for example: (1) performing the reaction with pH$_2$ at ultralow magnetic field to induce spontaneous polarization transfer~\cite{pravdivtsev2013c}; (2) applying an adiabatic magnetic field cycle~\cite{johannesson2004, goldman2005, cavallari2015} (FC), which is to perform the hydrogenation reaction at high field, nonadiabatically drop to ultralow field, and adiabatically return to high field, and; (3) applying an adiabatic magnetic field sweep~\cite{eills2019} (FS), which is to perform the hydrogenation at high field, then adiabatically reverse the magnetic field passing through zero field.

\begin{figure*}
\includegraphics[scale=1.4]{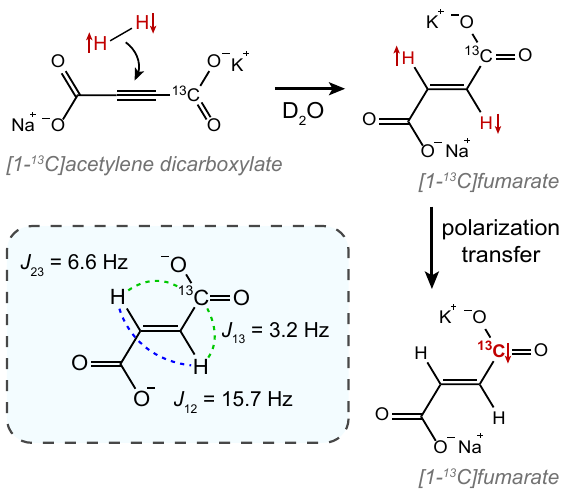}%
\caption{
\label{fig:reaction} 
The chemical reaction employed in this work to produce PHIP-polarized [1-\textsuperscript{13}C]fumarate. In the inset the molecule is shown with the J-couplings labelled.}
\end{figure*}

All NMR methods using adiabatic variation of the spin Hamiltonian are confronted with a common problem: adiabatic processes are by definition slow, and spin relaxation can be significant. Relaxation of hyperpolarized samples is generally detrimental as it gives rise to irreversible decay of the nonthermal spin order back to thermal equilibrium. It is therefore desirable to use the fastest possible adiabatic variation without disturbing the adiabatic nature of the process~\cite{messiah1962, morita2008, kaneko2015, roland2002}. Solutions have been proposed such as ``fast'' adiabatic processes given by optimal control theory~\cite{rodin2018} or by varying the Hamiltonian $\hat{H}(t)$ such that the effective adiabaticity parameter is constant at all times~\cite{rodin2019a}. The latter approach, constant-adiabaticity, is easy to implement and to adapt to specific molecular cases.

In this work we demonstrate constant-adiabaticity ultralow magnetic field manipulations to transfer proton singlet order into $^{13}$C magnetization in PHIP-polarized [1-\textsuperscript{13}C]fumarate. We form hyperpolarized fumarate by chemical reaction of para-enriched hydrogen with an acetylene dicarboxylate precursor molecule (see Fig.~\ref{fig:reaction}). The protons are initially in the singlet state, and are scalar-coupled to the $^{13}$C spin in the carboxylate position (we work at natural \textsuperscript{13}C abundance). In the case of [1-\textsuperscript{13}C]fumarate, the $J$-coupling between the protons is significantly larger than the proton-carbon $J$-couplings; this is referred to as the ``near-equivalence'' regime. As a consequence, the proton singlet state is close to an eigenstate, and significant state mixing which allows for polarization transfer occurs only at well-defined magnetic fields, $\pm B_{\text{LAC}}^{(i)}$, corresponding to the $i$-th level anti-crossings (LACs) of the spin system.~\cite{eills2019} Here we specifically investigate two ZULF methods to perform polarization transfer: field cycling, which uses a magnetic field variation from zero to $B_{\text{max}}$, and field sweeping which uses a magnetic field variation from $-B_{\text{max}}$ to $B_{\text{max}}$. For the case of [1-\textsuperscript{13}C]fumarate, $B_{\text{max}}$ is a few $\mu$T, which is considerably higher than the LAC fields, $B_{\text{LAC}}^{(i)}$. For both FC and FS experiments we derive constant-adiabaticity magnetic field profiles, $B(t)$, and compare the performance with linear (uniform) field variations.
\section{Theory}
\subsection{Hamiltonian}
The Hamiltonian of two protons, the $I$ spins ($I_1$ and $I_2$), and a $^{13}$C nucleus, the $S$ spin ($S_3$), in an external magnetic field $B(t)$ (aligned along the $z$-axis) is written as:
\begin{equation}
\label{eq:main_H}
    \hat{H}(t)=\hat{H}_Z(t)+\hat{H}_J\text{,}
\end{equation}
where
\begin{equation}
\label{eq:H_z}
    \hat{H}_Z(t)=-B(t)\{\gamma_I (\hat{I}_{1z}+\hat{I}_{2z})+\gamma_S \hat{S}_{3z}\}\text{,}
\end{equation}
\begin{equation}
\label{eq:H_JJ}
    \hat{H}_J=2 \pi J_{12} (\mathbf{\hat{I}}_1\cdot \mathbf{\hat{I}}_2)+
    2 \pi J_{13} (\mathbf{\hat{I}}_1\cdot\mathbf{\hat{S}})+
    2 \pi J_{23} (\mathbf{\hat{I}}_2\cdot\mathbf{\hat{S}}){,}
\end{equation}
and we set $\hbar=1$ for simplicity. At high magnetic field, and given that $J_{12}>|J_{13}-J_{23}|$, the eigestates of the Hamiltonian (\ref{eq:main_H}) are approximately equal to those of the $STZ$ (singlet-triplet-Zeeman) basis, which is defined as:
\begin{equation}
    \label{eq:STZ}
    STZ=\{ \ket{S^{12}},\ket{T_+^{12}},\ket{T_0^{12}},\ket{T_{-}^{12}}\} \otimes \{ \ket{\alpha_3},\ket{\beta_3} \}.
\end{equation}
The singlet and triplet states of the proton pair are defined as:
\begin{eqnarray}\label{eq:ST}
\ket{S^{12}} &=&(\ket{\a_1\b_2}-\ket{\b_1\a_2})/\sqrt{2},   \\\nonumber
\ket{T_{+1}^{12}} &=& \ket{\a_1\a_2},   \\\nonumber
\ket{T_0^{12}} &=& (\ket{\a_1\b_2}+\ket{\b_1\a_2})/\sqrt{2},   \\\nonumber
\ket{T_{-1}^{12}} &=& \ket{\b_1\b_2},
\end{eqnarray}
$\ket{\a}$ and $\ket{\b}$ denote the Zeeman spin states of an isolated spin-1/2 nucleus with $z$-projection of +1/2 and --1/2, respectively. The superscripts denoting the nucleus will be dropped henceforth. When the proton-carbon couplings are identical the eigenbasis is given exactly by eq. (\ref{eq:STZ}). However, when $J_{13}\neq J_{23}$, the protons are magnetically inequivalent which mixes the states, and the eigenbasis is then denoted $STZ'$. This is discussed in detail in Ref.~\onlinecite{eills2019}.

\begin{figure*}
\includegraphics[scale=0.5]{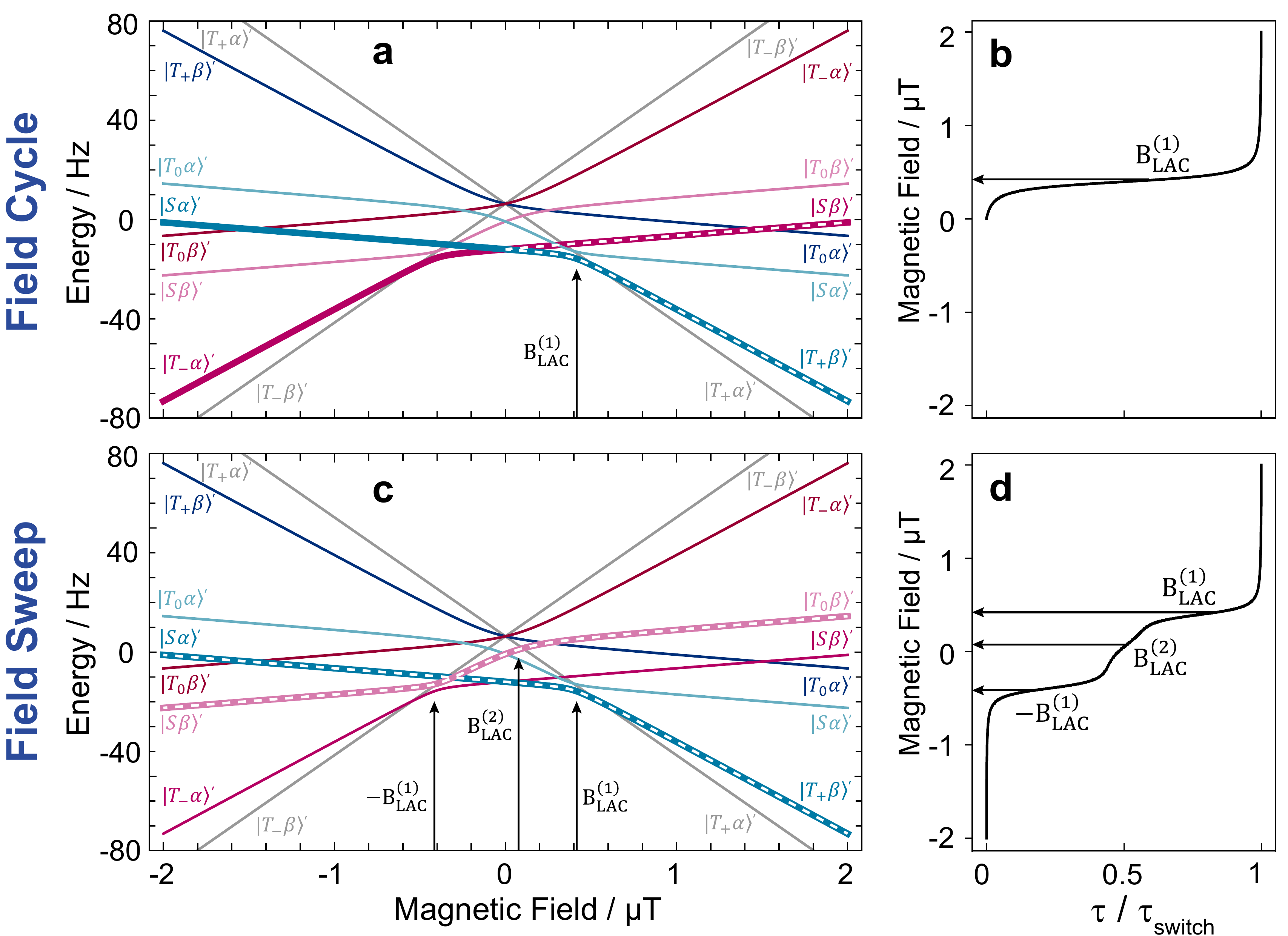}%
\caption{
\label{fig:en_levels_and_profiles} 
Eigenvalue plots for the 3-spin system of [1-\textsuperscript{13}C]fumarate with state labels shown. Top: in the eigenvalues plot in panel (a) the relevant energy levels for the field-cycling experiment are highlighted with white dashes, and the one relevant LAC is labelled. The constant-adiabaticity field-cycling profile is shown in panel (b).
Bottom: in the eigenvalues plot in panel (c) the relevant energy levels for the field-sweep experiment are highlighted, and the three relevant LACs are labelled. The constant-adiabaticity field-sweep profile is shown in panel (d).
For the constant-adiabaticity profiles, only the relevant (labelled) LACs are taken into account. The LAC magnetic field values for fumarate are $\pm B_{LAC}^{(1)}=\pm 0.416 \; \mu T$ and $\pm B_{LAC}^{(2)}=\pm 0.076 \; \mu T$.}
\end{figure*}

By plotting the eigenvalues of $STZ'$ as a function of magnetic field as shown in Fig.~\ref{fig:en_levels_and_profiles}, it can be seen that there are a number of energy level-crossings. On close inspection, one can see that in four places the crossings are in fact avoided; the inequivalence in proton-carbon couplings acts as a small perturbation which lifts the degeneracy of the crossing states, and the level crossings are turned into LACs. The positions of the LACs have been determined previously~\cite{eills2019}:
\begin{eqnarray}\label{eq:LACs}
    \pm B_{\text{LAC}}^{(1)}=\pm \frac{\pi(4J_{12}-(J_{13}+J_{23}))}{2(\gamma_I-\gamma_S)}, \\\nonumber
    \pm B_{\text{LAC}}^{(2)}=\pm \frac{\pi(J_{13}+J_{23})}{2(\gamma_I-\gamma_S)}.
\end{eqnarray}
At LACs, coherent spin mixing comes into play, which can be exploited in polarization-transfer experiments. For instance, adiabatic passage through a LAC gives rise to swapping of the populations of the unperturbed $STZ'$ states (commonly termed diabatic states). The reason is that the populations follow the instantaneous eigenstates, i.e., the eigenstates of the full Hamiltonian (defined in the presence of the perturbation terms). Hereafter, by adiabatic variation of the Hamiltonian $\hat{H}(t)$ of a spin system under study we mean that its eigenstates $\ket{\psi_i(t)}$ vary with time so slowly that the state populations have sufficient time to adjust to such changes (populations ``follow'' the time-dependent states). The focus of this work is to optimize adiabatic passage through LACs through the use of constant-adiabaticity field profiles to minimize the passage time.

\subsection{Density Matrix}
The nuclear spin state of $p$H$_2$ is given by the pure singlet-state wavefunction, since the two protons are magnetically equivalent. At magnetic fields far from $B_{LAC}^{(1)}$ the proton singlet state is also close to an eigenstate in [1-\textsuperscript{13}C]fumarate. Hence, when the hydrogenation reaction takes place at a magnetic field far from $B_{LAC}^{(1)}$ the singlet state remains close to an eigenstate. Note that the LACs at $\pm B_{\text{LAC}}^{(2)}$ occur between the proton triplet states, so the hydrogenation could be performed at zero-field and the proton singlet population still substantially retained. The initial density matrix is approximately:
\begin{equation}
    \rho_0\approx\frac{1}{2}\left\{\ket{S\alpha}'\bra{ S\alpha}'+\ket{S\beta}'\bra{ S\beta}'\right\},
\end{equation}
where the primes indicate that the eigenstates are from the $STZ'$ basis, not $STZ$. Only the two states close to singlet states of the two protons are populated; since $J_{12}\gg |J_{13}-J_{23}|$ the populations of other six spin states are negligibly small.

\subsection{Constant adiabaticity profile}
The general adiabaticity parameter is defined as~\cite{rodin2019a}:
\begin{eqnarray}\label{eq:GAP}
    \xi(t)&=&\sqrt{\sum_{i,j}|\xi_{ij}|^2},~~~\\\nonumber&&{\rm where}~~\xi_{ij}=\frac{\bra{i} \frac{d}{dt} \ket{j}}{\omega_{ij}}=\frac{\bra{i}\frac{d\hat{H}}{dt} \ket{j}}{\omega_{ij}^2},
\end{eqnarray}
where $\ket{i},\ket{j}$ are the eigenstates of the Hamiltonian (\ref{eq:main_H}) and $\omega_{ij}=\omega_i-\omega_j$ is their energy difference (expressed in angular frequency units). Here the Haniltonian derivative is much easier to calculate then the eigenstate derivative. Hence, as dictated by eq. (\ref{eq:GAP}), for each pair of states we need to compute the parameter $\xi_{ij}$,  which defines how fast the eigenstates change with time compared to the internal evolution frequency given by $\omega_{ij}$. After that, we evaluate the general adiabaticity parameter $\xi$ by averaging over $\xi_{ij}$ defined for each pair of states. When $\xi\ll 1$, the process is adiabatic and the populations remain in the instantaneous eigenstates.

In order to determine optimized $B(t)$ ramps, we impose the condition that the  general adiabaticity parameter is equal to a constant value, $\xi(t)=\xi_0$, during the variation. Before proceeding, we introduce a few improvements for calculating constant-adiabaticity $B(t)$ profiles.

First of all, to transfer populations between the diabatic states we use LACs, but there are also many level crossings in the system which occur between two states of degenerate energy when there are no perturbation terms to induce state mixing. Level crossings of a pair of levels occur when the Hamiltonian has a block-diagonal structure and the two states belong to different blocks. In the case under study, the block-diagonal structure of the Hamiltonian is dictated by the fact that the $z$-projection of the total spin is conserved, since the commutator $[\hat{H},\hat{I}_{1z}+\hat{I}_{2z}+\hat{S}_z]$ is zero. We exclude level crossings from consideration since, as follows from Eq. (\ref{eq:GAP}), calculation of the $\bra{i}\xi\ket{j}$ parameter meets certain difficulties (the numerator and denominator tend to zero). Although this uncertainty in calculating $\xi_{ij}$ can be resolved analytically, numerical calculation of the adiabaticity parameter becomes problematic. Hence, we need to evaluate the $\xi_{ij}$ parameters only for the states belonging to the same blocks and Eq. (\ref{eq:GAP}) can be modified as follows:
\begin{equation}\label{eq:GAP_blocks}
    \xi(t)=\sqrt{
        \sum_{i^{(p)},j^{(p)}}  
        |\xi_{i^{(p)},j^{(p)}}|^2
        },
\end{equation}
where $\ket{i^{(p)}},\ket{j^{(p)}}$ are the eigenstates of the Hamiltonian (\ref{eq:main_H}) belonging to the block or subspace $p$. In our particular case, the index $p$ refers to the angular momentum projection $m$ which equals $m=\pm 3/2,~\pm 1/2$.

Second, we take into account that in the case of [1-\textsuperscript{13}C]fumarate prepared via PHIP, to a good approximation only the $\ket{S\alpha}'$ and $\ket{S\beta}'$ states are populated, which is only two states out of eight. To adiabatically manipulate the spin order, it is sufficient to consider only mixing of these states with other states belonging to the same blocks in the Hamiltonian. Specifically, the block of spin states characterized by the angular momentum projection on the field axis $m=1/2$ comprises three states $\ket{T_{0}\alpha}'$, $\ket{S\alpha}'$ and $\ket{T_{+}\beta}'$. If initially only the $\ket{S\alpha}'$ state is populated, spin mixing only in pairs of states $\ket{S\alpha}'\leftrightarrow\ket{T_0\alpha}'$ and $\ket{S\alpha}'\leftrightarrow\ket{T_{+}\beta}'$ is important.
To take this into account, Eq. \ref{eq:GAP_blocks} should be modified as follows:
\begin{equation}\label{eq:GAP_final}
    \xi(t)=\sqrt{
        \sum_{i^{(p)}_0,j^{(p)}}  
        \left|
        \xi_{
        i^{(p)}_0, j^{(p)}
        }
        \right|^2
        }\text{,}
\end{equation}
where $\ket{i^{(p)}_0}$ are the spin states belonging to subspace $p$ with non-zero initial population, due to the chosen method of preparing the system.

With these two considerations, we can calculate the optimized $B(t)$ profile. The time derivative of the Hamiltonian (\ref{eq:main_H}) is:
\begin{equation}\label{eq:H_der}
    \frac{d\hat{H}}{dt}=-\frac{dB}{dt}(\gamma_I (\hat{I}_{1z}+\hat{I}_{2z})+\gamma_S \hat{S}_{3z})\text{.}
\end{equation}
By substituting (\ref{eq:H_der}) into (\ref{eq:GAP_final}) and setting the general adiabaticity parameter to a constant value $\xi(t)=\xi_0$, we obtain:
\begin{equation}\label{eq:profile}
    \frac{dB}{dt}=
    \xi_0   \left/
    \sqrt{\sum_{i_0^{(m)},j^{(m)}} 
    \frac{
    \left|\bra{i_0^{(m)}} \gamma_I \hat{I}_z+\gamma_S \hat{S}_{3z} \ket{j^{(m)}_{\phantom{0}}}
    \right|^2
    }{
    2\omega_{i_0^{(m)},j^{(m)}}^4}
    }\right. \text{,}
\end{equation}
here $\hat{I}_z=\hat{I}_{1z}+\hat{I}_{2z}$. By integrating this expression numerically, we calculate the sought optimized ``constant-adiabaticity'' $B(t)$ ramp. 

\subsection{Polarization-transfer methods}
The idea of the two methods considered here, FC and FS, is illustrated by the energy level diagrams shown in Figure \ref{fig:en_levels_and_profiles}, which highlight the polarization-transfer pathways. In both experiments, the hydrogenation step is performed at +2~$\mu$T to produce [1-\textsuperscript{13}C]fumarate with the $|S\alpha\rangle'$ and $|S\beta\rangle'$ states populated.

In the FC experiment, the field is then rapidly (nonadiabatically) dropped to zero, which preserves the populations of $|S\alpha\rangle'$ and $|S\beta\rangle'$, and then abiabatically increased to exchange the populations of $|S\alpha\rangle'$ and $|T_+\beta\rangle'$, but leave the population of $|S\beta\rangle'$ unchanged. At the end of the field cycle the $|T_+\beta\rangle'$ and $|S\beta\rangle'$ states are populated, and hence the \textsuperscript{13}C spin is hyperpolarized. In this experiment only one LAC is relevant: the LAC at $B_{\text{LAC}}^{(1)}$.

In the FS experiment, after the hydrogenation step the field is reversed rapidly (nonadiabatically) to --2~$\mu$T (although the hydrogenation could be done at --2~$\mu$T and this step skipped) which preserves the populations of $|S\alpha\rangle'$ and $|S\beta\rangle'$, and then increased adiabatically through zero to +2~$\mu$T. The population in $|S\alpha\rangle'$ ends in $|T_{+}\beta\rangle'$, and the population in $|S\beta\rangle'$ ends in $|T_{0}\beta\rangle'$. At the end of the field sweep, the $|T_{+}\beta\rangle'$ and $|T_{0}\beta\rangle'$ states are populated, and hence the \textsuperscript{13}C spin is hyperpolarized. In this experiment three LACs are relevant: LACs occurring at the fields $-B_{\text{LAC}}^{(1)}$, $B_{\text{LAC}}^{(1)}$ and $B_{\text{LAC}}^{(2)}$. Here we want to stress here that because $-B_{\text{LAC}}^{(2)}$ is not part of the adiabatic pathway, the resulting field profile is slightly asymmetric with respect the center point.

In Figure \ref{fig:en_levels_and_profiles}, we also show the constant adiabaticity $B(t)$ ramps for both cases. One can see that the proposed algorithm dictates a slow increase of the field at the LACs, whereas away from the LACs switching can be done fast.

\section{Materials and Methods}
All chemicals were purchased from Sigma Aldrich. A solution of 50 mM monopotassium acetylene dicarboxylate, 100 mM sodium sulphite and 7 mM ruthenium catalyst [RuCp*(CH\textsubscript{3}CN)\textsubscript{3}]PF\textsubscript{6} in D\textsubscript{2}O was prepared by dissolving the solids by heating and sonication. The sodium sulphite was added to increase the rate of reaction as discussed in Refs.~[\cite{ripka_hyperpolarized_2018,knecht2020}]. The pH of the solution was adjusted to pH 10 with NaOD to further improve the rate of reaction. Oxygen was removed from the solution by bubbling nitrogen through for 5 minutes. 300 \microL of this precursor solution was used for each experiment.

The NMR experiments were performed in a 1.4 T \textsuperscript{1}H-\textsuperscript{13}C dual resonance SpinSolve NMR system (Magritek, Aachen).

Parahydrogen at $>$98\% enrichment was generated by passing hydrogen gas ($>$99.999\% purity) through an Advanced Research Systems (ARS, Macungie, USA) parahydrogen generator operating at 25 K.

For ultralow-field experiments, a magnetic shield (MS-1F, Twinleaf LLC, Princeton, USA) was used to provide a 10\textsuperscript{6} shielding factor against external magnetic fields. Static internal magnetic fields for shimming were produced using built-in B\textsubscript{x}, B\textsubscript{y}, and B\textsubscript{z} coils, powered with computer-controlled DC calibrators (Krohn-Hite, model 523, Brockton, USA), providing three-axis field control. The time dependent applied magnetic fields were generated with a Helmholtz coil (70 mm diameter) wound on a 3D-printed former, with current supplied by a power amplifier (AE Techron 7224-P, Elkhart, USA). The magnetic-field profiles were generated using a data acquisition card (NI-9263, National Instruments, Austin, USA) with 10 $\mu$s time precision.
\begin{figure}
\includegraphics[width=\linewidth]{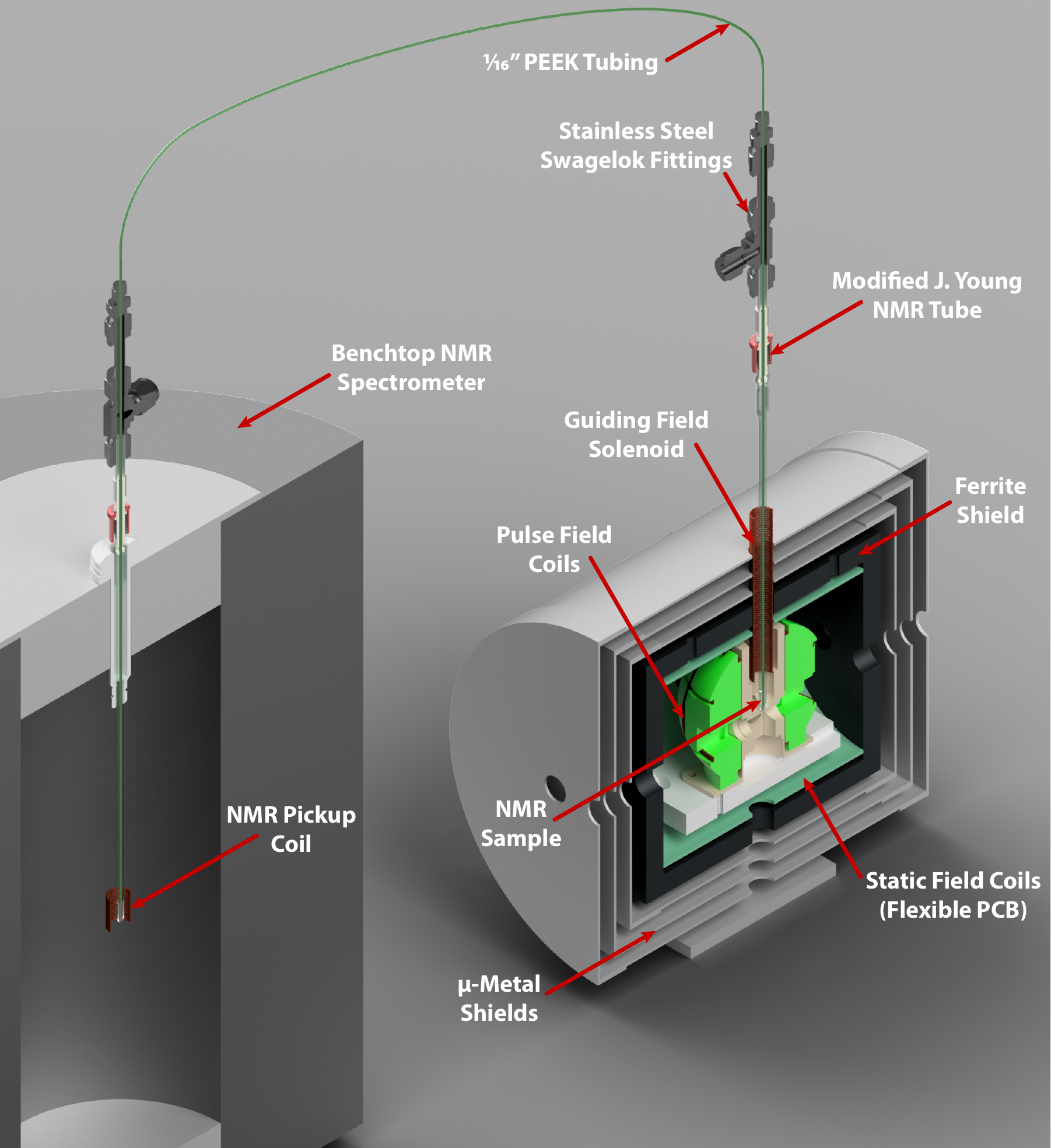}%
\caption{
\label{fig:schematic} 
The experimental apparatus used in this work.
Image reproduced from Ref.~\onlinecite{eills2019} with the permission of AIP Publishing.}
\end{figure}

Low-pressure/vacuum J. Young NMR tubes held in the ZULF chamber and 1.4 T SpinSolve NMR spectrometer were connected with polytetrafluoroethylene (PTFE) tubing (1/16 in. O.D., 0.5 mm I.D.), as shown in Fig.~\ref{fig:schematic}. Gas and liquid flow were controlled by pneumatically actuated valves (Swagelok, Frankfurt, Germany). Sample hydrogenation was followed by shuttling into the SpinSolve by reversing the gas flow. The sample transport was performed with nitrogen gas (any unreactive gas could be used) and took 2.5\,s. In order to prevent the sample from passing through any fields that could lead to undesired state-mixing during sample transport, a penetrating solenoid was used to provide a guiding field during transit out of the magnetic shield. To avoid having bubbles in the detection region after sample transport, 100 \microL of acetone was placed in the SpinSolve tube at the start of each experiment. This mixed with the fumarate solution after shuttling, and served to reduce the surface tension and viscosity of the D\textsubscript{2}O solvent. The experimental apparatus is shown in Fig.~\ref{fig:schematic}.

At the start of the experiment the sample was in the ZULF chamber in a 5 mm NMR tube, in a +2 \microT (chosen as a relatively low field that is still high enough for the Hamiltonian eigenstates to be, to a good approximation, the $STZ'$ basis states) field provided by the Helmholtz coils, and parahydrogen gas was bubbled in at 7 bar for 30 s. After a 1 s delay to allow the sample to settle, a field manipulation was applied using the Helmholtz coils. After the field sweep/cycle, the solenoid guiding field was switched on to provide a +20 \microT field, and nitrogen gas at 7 bar was used to shuttle the sample into the SpinSolve NMR spectrometer. After a 1 s delay for the sample to settle, a 90$\deg$ pulse was applied followed by data acquisition. A simplified event sequence is shown in Fig.~\ref{fig:exp_and_protocol}.

After the hyperpolarization had fully relaxed, a thermal equilibrium \textsuperscript{1}H NMR spectrum was acquired on each sample to quantify the concentration of fumarate formed. The hyperpolarized \textsuperscript{13}C NMR results were normalized against this, to account for the differences in reaction yield between experiments.

\section{Results and Discussion}

\begin{figure*}[th]
\includegraphics[scale=1.2]{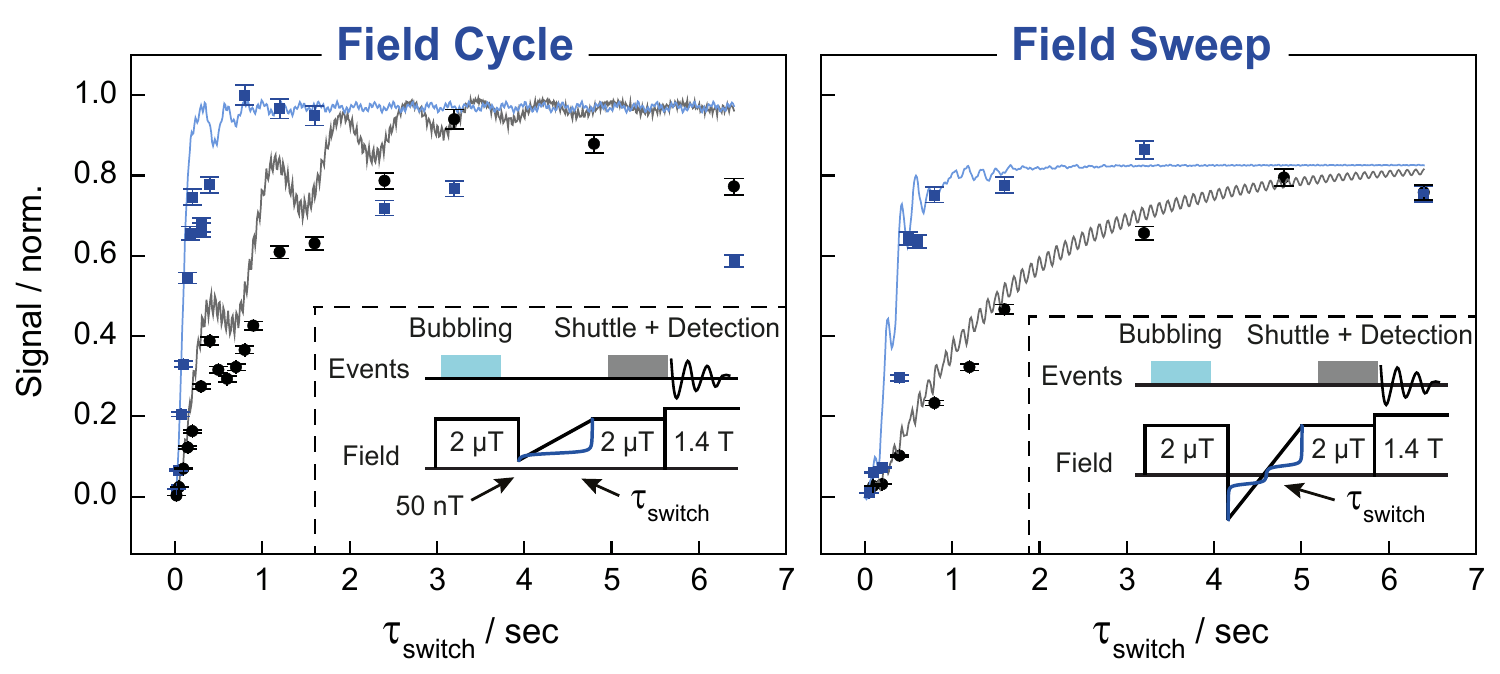}%
\caption{
\label{fig:exp_and_protocol} 
Results from the field cycling and field sweeping experiments, contrasting constant-adiabaticity and linear magnetic field profiles. The error bars on the data points show one standard deviation, determined by repeating one experiment many times and measuring the standard deviation. The data points are normalized to 1, corresponding to the highest-observed signal. Simulations are shown by the lines fitted to the data. A description of how the simulations were performed is given in the text. The experimental protocols are shown in the insets. In theory and in simulations, both methods can lead to $>$97\% \textsuperscript{13}C polarization. The lower efficiency of the field sweep experiment is discussed in the text.}
\end{figure*}

The results from experiments comparing linear and constant-adiabaticity field cycling and field sweeping are shown in Fig.~\ref{fig:exp_and_protocol}. Each data point represents the signal from one experimental run. Simulations of the transfer efficiency as a function of sweep duration are shown by the lines (which ignore relaxation effects). In both cases, using the constant-adiabaticity profile allows one to achieve the maximal $^{13}$C polarization faster than by using a linear profile. The faster spin-order conversion helps to minimize loss of polarization due to relaxation, although in these particular experiments the observed \textsuperscript{13}C polarization is similar between the constant-adiabaticity and linear experiments. This is because the spin relaxation times are relatively long compared to the duration of the magnetic field manipulations, and significant polarization loss is only observed for long switching times. The overall switching times are shorter for the FC experiment which requires passage through only one LAC. Note that the nonadiabatic field reversal at the start of the field sweep experiment was used for convenience, but isn't expected to have any effect on the spin dynamics; the hydrogenation could equally be performed at $B=-2\,\mu$T and the field adiabatically increased from there.

To perform the spin dynamics simulations, firstly the density matrix is projected onto the eigenbasis of the Hamiltonian (\ref{eq:main_H}) defined at $B=2$ $\mu$T and all off-diagonal elements (coherences) of the density matrix are removed, since they are averaged out upon continuous production of polarized molecule by the hydrogenation reaction. The resulting density matrix describes the spin system immediately following the hydrogenation step. Next, we numerically solve the Liouville-von Neumann equation with the time-dependent Hamiltonian (\ref{eq:main_H}), where $B(t)$ corresponds to the magnetic field profile. Finally, we extract the expectation value of $S$-spin polarization from the final density matrix. Relaxation was not included in the simulations.

\begin{figure*}
\includegraphics[width=\linewidth]{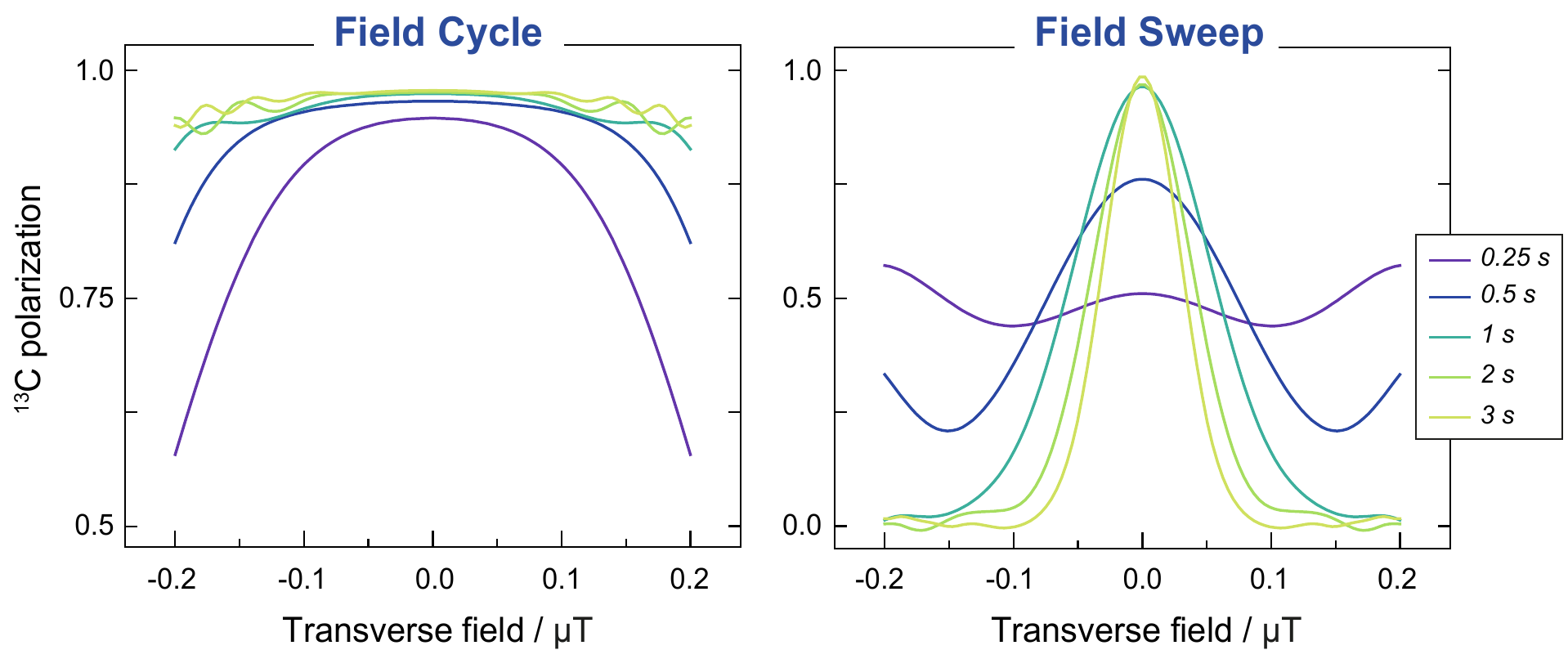}%
\caption{
\label{fig:Fig5} 
The \textsuperscript{13}C polarization generated by applying a constant-adiabaticity field cycle (left) or constant-adiabaticity field sweep (right) to [1-\textsuperscript{13}C]fumarate for different static transverse (i.e. in the $x/y$-plane) magnetic field strengths. In the simulated field cycle $B_z$ is cycled from 50\,nT to 2\,$\mu$T and in the simulated field sweep $B_z$ is swept from -2 to +2\,$\mu$T. The field cycle is less sensitive to static transverse fields since the level anti-crossings introduced by the transverse field are centered at $B_z=0$. Simulations were performed for a number of field cycle/field sweep durations.
}
\end{figure*}

The experimental results are generally in good agreement with the simulations, but the \textsuperscript{13}C signals for the field sweeping experiments are notably lower than in the field cycling experiments, with field cycling showing $\sim$15\% higher transfer efficiency. This is not intrinsic to the methodology, since both methods can lead to $>97\%$ \textsuperscript{13}C polarization in this molecular system. We believe the lower efficiency of the field sweep is predominantly for two reasons. Firstly, in the field sweep experiment, adiabatic passage through three LACs is necessary for polarization transfer, whereas for the field cycling experiment only one LAC is used. Imperfections in the adiabatic passages will therefore compound, and be more detrimental in the field sweep experiment. Secondly, and likely more importantly for most experimental cases, the requirement to pass through zero magnetic field for the field sweep experiment can cause significant loss of polarization if there are residual magnetic fields along other axes. This introduces additional undesirable LACs which can lead to the populations being diverted from the desired transfer path. In Fig.\,\ref{fig:Fig5} we show how the final \textsuperscript{13}C polarization for a magnetic field cycle/sweep in the $z$-axis depends on the presence of a transverse field in the $x/y$-plane. We now use $B_z$ to indicate a $B$ field applied along the $z$-axis.


\begin{figure*}
\includegraphics[width=\linewidth]{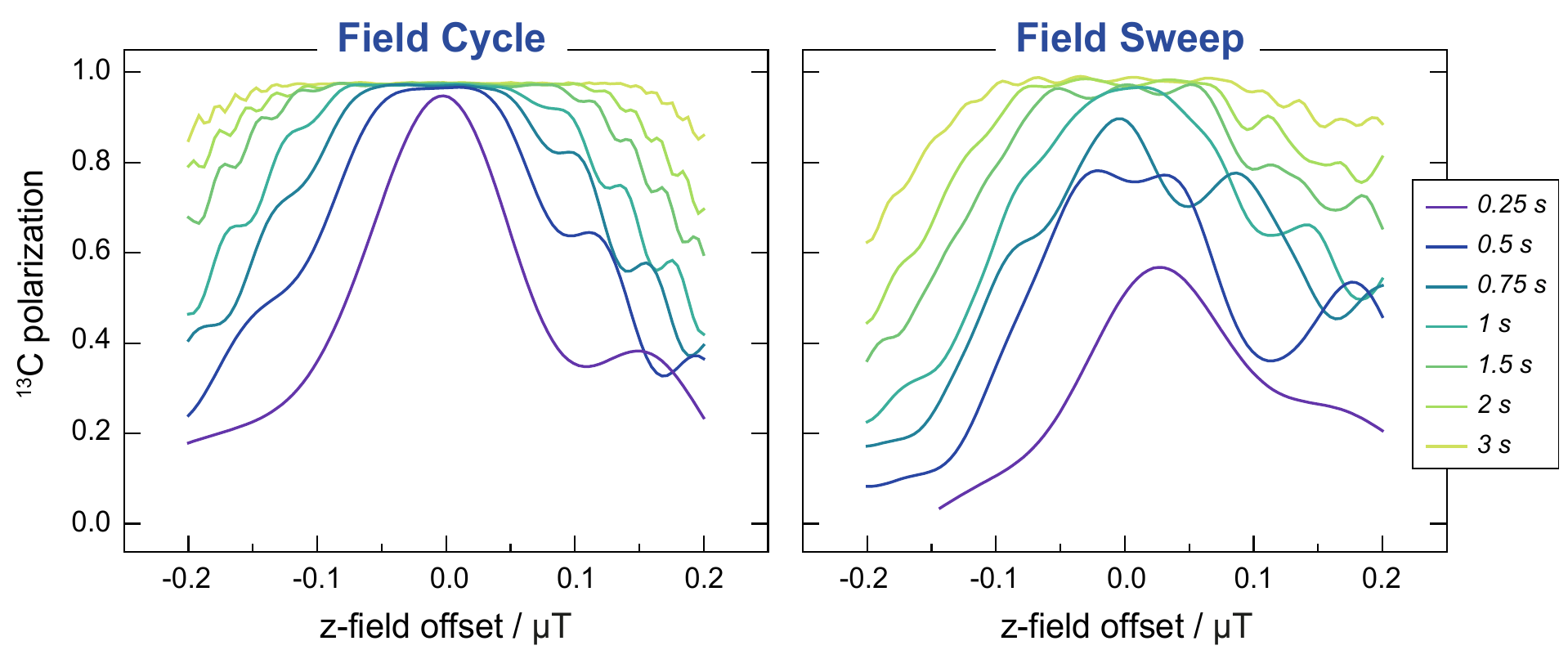}%
\caption{
\label{fig:Fig6} 
The \textsuperscript{13}C polarization generated by applying a constant-adiabaticity field cycle (left) or constant-adiabaticity field sweep (right) to [1-\textsuperscript{13}C]fumarate, as a function of $B_z$ magnetic field offset. In the simulated field cycle $B_z$ is cycled from 50\,nT to 2\,$\mu$T and in the simulated field sweep $B_z$ is swept from -2 to +2\,$\mu$T. Simulations were performed for a number of field cycle/field sweep durations, and the lines are labelled correspondingly.
}
\end{figure*}

Despite shorter switching times when using a constant-adiabaticity profile, these methods are more sensitive to magnetic field offset in the field sweep axis ($z$) than the linear profiles. This is because the constant-adiabaticity profiles are designed around the knowledge of LAC fields, and if there is a magnetic field offset or inhomogeneity across the sample, the slow part of the constant-adiabaticity field ramp will not match the LAC field. The dependence of the constant-adiabaticity FC and FS conversion efficiencies on magnetic field offset is shown in Fig.~\ref{fig:Fig6}. When using optimized FC/FS parameters, a $B_z$ offset on the order of 100~nT is sufficient to reduce the transfer efficiency by $>$10\%. The case is worse for the FS experiment, which requires three LACs, compared to just one for the FC experiment. We expect this situation can be improved by designing pseudo-constant-adiabaticity profiles to be close to constant-adiabaticity, but made to be more robust with respect to $B_z$ field offset/inhomogeneity by broadening the LAC field condition.

In these experiments the field cycling experiment performs better than field sweeping. However, there are certainly experimental cases in which the field sweep might be preferred. One example is in a system in which the rapid (nonadiabatic) initial field drop of the field cycle is inconvenient or not possible, such as when instead of varying the magnetic field applied to a static liquid, a static magnetic field spatial profile is constructed and the liquid flows through to achieve the desired polarization transfer. Another example of when the field sweep experiment may be preferred is for molecular systems outside the near-equivalence regime (i.e. when $|J_{12}|<|\delta J|$, in our case $J_{12}=15.7$ Hz and $\delta J = 3.4$ Hz). In these cases, the proton singlet state is no longer close to an eigenstate at zero field, and the nonadiabatic field switch at the start of the field-cycling experiment can cause the population differences to be converted into coherences which rapidly dephase. This can be avoided by using a magnetic field sweep.

\section{Conclusions}
In this work we revisited the concept of using adiabatic variation of the spin Hamiltonian for manipulating nonthermal spin order. By exploiting the constraint of ``constant adiabaticity'' we are able to increase the rate of spin-order transformations in nuclear spin systems. In addition to the previously developed algorithm, here we propose a modification of the method, allowing one to limit the number of adiabatic levels to only those relevant for the desired spin-order transfer. The theoretical approaches discussed here are of a general scope, and they can be applied to a variety of NMR (and non-NMR) experiments. We illustrate the performance of constant-adiabaticity optimization on the specific example of polarization transfer from parahydrogen-derived proton singlet order to a heteronuclear magnetization in an AA$'$X spin system in ZULF conditions. In such experiments the external magnetic field is adiabatically varied in the $\mu$T range; specifically, it is swept through zero field or cycled between zero field and a field on the order of $\mu$T. We demonstrate the method on the molecule [1-\textsuperscript{13}C]fumarate, and show that constant-adiabaticity $B(t)$ ramps provide faster spin-order transfer than linear ramps, which is important when detrimental relaxation effects are considered. We expect that variation of the Hamiltonian using the constraint of constant adiabaticity will become a useful tool in NMR in general, and in ZULF NMR in particular.

\begin{acknowledgments}
The authors would like to acknowledge Dr. John W. Blanchard and Dr. Teng Wu for experimental advice. This research was supported the Russian Science Foundation (project 20-62-47038). This project has received funding from the European Union's Horizon 2020 research and innovation programme under the Marie Skłodowska‐Curie Grant Agreement No. 766402.
\end{acknowledgments}

\section*{References}
\bibliography{lib_auth_year_no_notes2}
\end{document}

%% file: symbols/main.tex
\input symbols/MHL-math.tex

\input symbols/MHL-text.tex
\input symbols/revisions.tex

%% file: symbols/MHL-math.tex
\newcommand\abs[1]{\left|#1\right|}
\newcommand\norm[1]{\left\|#1\right\|}
\newcommand\btheta{\beta_\theta}
\newcommand\One{{\mathbbm 1}}
\newcommand\Bfac{{\mathbbm B}}
\newcommand{\kB}{k_B}
\newcommand\hf{\tfrac{1}{2}}
\newcommand{\inv}{^{-1}}
\renewcommand{\deg}{^\circ}
\renewcommand\a{\alpha}
\renewcommand\b{\beta}
\newcommand\g{\gamma}
\renewcommand\d{\delta}
\newcommand{\rhoeq}{\rho_\mathrm{eq}}
\newcommand{\Seq}{S_\mathrm{eq}}
\newcommand{\Ix}{I_x}
\newcommand{\Iy}{I_y}
\newcommand{\Iz}{I_z}
\newcommand{\Tone}{T_1}
\newcommand{\TS}{T_S}
\newcommand{\ta}{t_a}
\newcommand{\tb}{t_b}
\newcommand{\tauev}{\tau_{\rm ev}}
\newcommand{\tauT}{\tau_T}
\newcommand{\tauopt}{\tau_{\rm opt}}
\newcommand{\Rone}{R_1}
\newcommand{\RS}{R_S}
\newcommand{\Bzero}{B^0}
\newcommand{\wzero}{\omega^0}
\newcommand{\NH}{N_H}
\newcommand\ket[1]{\left\vert#1\right>}
\newcommand\bra[1]{\left<#1\right\vert}
\newcommand{\UAPSOC}{U_\mathrm{APSOC}}
\newcommand\NL{N_{\mathcal{L}}}
\newcommand\Ket[1]{\left\vert#1\right)}
\newcommand\Bra[1]{\left(#1\right\vert}
\newcommand\BraKet[2]{\left(#1\big{|}#2\right)}
\newcommand\Tr[1]{\mathrm{Tr}\!\left\{#1\right\}}
\newcommand\Evals{{\mathbf{\Lambda}}}
\newcommand{\QZ}{Q_Z}
\newcommand{\QS}{Q_S}
\newcommand{\Qtheta}{Q_\theta}
\newcommand{\B}{\mathcal{B}}
\newcommand{\rB}{r_\mathcal{B}}
\newcommand{\pvec}{\mathbf{p}}
\newcommand{\pveceq}{\mathbf{p}_\mathrm{eq}}
\newcommand{\pvecss}{\mathbf{p}_\mathrm{ss}}
\newcommand{\pvecfin}{\mathbf{p}_\mathrm{fin}}
\newcommand{\Zvec}{\mathbf{Z}}
\newcommand{\Svec}{\mathbf{S}}
\newcommand{\ZO}{\langle Z \rangle}
\newcommand{\ZOeq}{\ZO_\mathrm{eq}}
\newcommand{\ZOss}{\ZO_\mathrm{ss}}
\newcommand{\ZOfin}{\ZO_\mathrm{fin}}
\newcommand{\SO}{\langle S \rangle}
\newcommand{\SOeq}{\SO_\mathrm{eq}}
\newcommand{\SOss}{\SO_\mathrm{ss}}
\newcommand\pimat[1]{\bm{\pi}_{#1}}
\newcommand{\Cmat}{\mathbf{C}}
\newcommand\ThTreset{\bm{\Theta}_T^\mathrm{reset}}
\newcommand{\rhoZ}{\rho_Z}
\newcommand{\rhoS}{\rho_S}
\newcommand{\rhoZeq}{\rho_Z^\mathrm{eq}}
\newcommand{\rhoSeq}{\rho_S^\mathrm{eq}}
\newcommand{\rhoZss}{\rho_Z^\mathrm{ss}}
\newcommand{\rhoSss}{\rho_S^\mathrm{ss}}
\newcommand{\rhoSUmax}{\rho_S^{U,\mathrm{max}}}
\newcommand{\rhoZfinal}{\rho_Z^{\mathrm{final}}}
\newcommand{\hatU}{\hat{U}}
\newcommand{\hatV}{\hat{V}}
\newcommand{\sopG}{{\hat\Gamma}}
\newcommand{\sopT}[2]{\hat{T}_{#2}^{#1}}
\newcommand{\Cseq}{\mathcal{C}}
\newcommand{\Sseq}{\mathcal{S}}
\newcommand{\nC}{n_C}
\newcommand{\nP}{n_\mathcal{P}}
\newcommand{\Tzz}{T$_{00}$\xspace}
\newcommand{\wAPSOC}{\omega_{\rm APSOC}}
\newcommand{\wAPSOCmax}{\wAPSOC^{\rm max}}
\newcommand{\TAPSOC}{T_{\rm APSOC}}
\newcommand{\Dd}{\Delta\delta}
\newcommand{\wD}{\omega_\Delta}
\newcommand{\aMSM}{a_{\rm MSM}}
\newcommand{\aMSMU}{\aMSM^{U}}
\newcommand{\aMSMUmax}{\aMSM^{U,\rm max}}
\newcommand{\wran}{\omega_{\rm ran}}
\newcommand{\tauc}{\tau_c}
\newcommand{\Wth}{\boldsymbol{W}^\theta}
\newcommand{\Vth}{\boldsymbol{V}^\theta}
\newcommand{\popeq}{\boldsymbol{p}^{\rm eq}}
\newcommand{\popss}{\boldsymbol{p}^{\rm ss}}
\newcommand{\Pimat}{\boldsymbol{\Pi}}

%% file: symbols/MHL-text.tex
\newcommand\blue[1]{\textcolor{blue}{#1}}
\newcommand\red[1]{\textcolor{red}{#1}}
\newcommand\Comment[1]{\blue{\tt [#1]}}
\newcommand\ToHere{\ \newline\Comment{*** MHL edits up to here ***}\ \newline}
%
\newcommand\KInote[1]{\textcolor{orange}{[KI: #1]}}
\newcommand{\BRnote}[1]{\textcolor{cyan}{[BR: #1]}}
\newcommand\Schrodinger{Schr{\"o}dinger\xspace}
\newcommand\Sorensen{S{\o}rensen\xspace}
\newcommand\ortho{{\em ortho}\xspace}
\newcommand\para{{\em para}\xspace}
\newcommand\SpinDynamica{{\em SpinDynamica}\xspace}
\newcommand{\Cth}{$^{13}{\rm C}$\xspace}
\newcommand{\Ctwo}{$^{13}{\rm C}_2$\xspace}
\newcommand{\prot}{$^{1}{\rm H}$\xspace}
\newcommand{\CHthree}{${\rm CH}_3$\xspace}
\newcommand{\CtwoOne}{$^{13}{\rm C}_2$-\textbf{I}\xspace}
\newcommand{\acetoned}[1]{acetone-d$_#1$}
\newcommand{\microL}{$\mu$L\xspace}
\newcommand{\microT}{$\mu$T\xspace}

%% file: symbols/revisions.tex
\newcommand\revision[1]{{#1}}

%% file: ReferenceSets.tex
\def\citeACrefs/{\cite{schulman_molecular_1999,fernandez_algorithmic_2004-1,sklarz_optimal_2004,schulman_physical_2005,baugh_experimental_2005,ryan_spin_2008,elias_heat-bath_2011,yuan_reachable_2013,brassard_experimental_2014,raeisi_asymptotic_2015,li_approximation_2016,rodriguez-briones_achievable_2016,rodriguez-briones_heat-bath_2017}}
\def\refsAC{%
boykin_algorithmic_2002,%
fernandez_algorithmic_2004,%
sklarz_optimal_2004,%
schulman_physical_2005,%
yuan_reachable_2013,%
raeisi_asymptotic_2015,%
li_approximation_2016,%
rodriguez-briones_achievable_2016,%
rodriguez-briones_heat-bath_2017,%
baugh_experimental_2005,%
fernandez_paramagnetic_2005,%
moussa_heat-bath_2005,%
ryan_spin_2008,%
elias_heat-bath_2011,%
jones_NMRQCompReview_2011,%
brassard_experimental_2014,%
khurana_bang-bang_2017%
}
\def\refsACNMRexpt{%
baugh_experimental_2005,%
fernandez_paramagnetic_2005,%
moussa_heat-bath_2005,%
ryan_spin_2008,%
elias_heat-bath_2011,%
jones_NMRQCompReview_2011,%
brassard_experimental_2014,%
khurana_bang-bang_2017%
}
\def\refsACbounds{%
schulman_physical_2005,%
yuan_reachable_2013,%
raeisi_asymptotic_2015,%
li_approximation_2016,%
rodriguez-briones_achievable_2016%
}
\def\refsSingletNMR{%
carravetta_beyond_2004,%
carravetta_long-lived_2004,%
sarkar_singlet-state_2007,%
pileio_long-lived_2008,%
warren_increasing_2009,%
pileio_relaxation_2010,%
pileio_storage_2010,%
tayler_singlet_2011,%
levitt_singlet_2012,%
tayler_accessing_2013,%
stevanato_nuclear_2015,%
zhang_singlet_2015,%
zhang_limits_2016,%
kharkov_effect_2017,%
pravdivtsev_robust_2016,%
rodin_using_2018,%
sheberstov_cis_2018,%
rodin_constant-adiabaticity_2019,%
kharkov_singlet_2019,%
kiryutin_proton_2019,%
levitt_long_2019%
}
\def\refsAPSOC{%
pravdivtsev_robust_2016,%
rodin_using_2018,%
sheberstov_cis_2018,%
rodin_constant-adiabaticity_2019%
}
\def\refsSTop{%
wokaun_selective_1977,%
vega_fictitious_1978%
}
\def\refsCompPulse{%
levitt_nmr_1979,%
levitt_SymmCompPulseRF_1982,%
levitt_composite_1986,%
wimperis_broadband_1994,%
vandersypen_NMRQuantumControlReview_2005%
}
%
\def\refsUbounds{%
sorensen_polarization_1989,%
sorensen_universal_1990,%
sorensen_entropy_1991,%
levitt_unitary_1992,%
nielsen_bounds_1995,%
stoustrup_generalized_1995,%
levitt_symmetry_2016%
}
\def\refsNMRQuantumComputing{%
cory_NMR-QComp_1997,%
vandersypen_NMRQuantumControlReview_2005,%
jones_NMRQCompReview_2011%
}